\documentclass[aip,reprint,amsmath,amssymb,floatfix]{revtex4-1}

\usepackage[utf8]{inputenc}
\usepackage[T1]{fontenc}

\usepackage{graphicx}
\usepackage[dvipsnames]{xcolor}
\usepackage{bm}
\usepackage{mathptmx}
\usepackage{etoolbox}

\begin{document}
\title{Viscosity of polymer melts using non-affine theory based on vibrational modes}

\author{Ankit Singh}
\email{ankit.singh@unimi.it}
\affiliation{Department of Physics ``A. Pontremoli", University of Milan, via Celoria 16, 20133 Milan, Italy}

\author{Vinay Vaibhav}
\email{vinayphys@gmail.com}
\affiliation{Department of Physics ``A. Pontremoli", University of Milan, via Celoria 16, 20133 Milan, Italy}


\author{Alessio Zaccone}
\email{alessio.zaccone@unimi.it}
\affiliation{Department of Physics ``A. Pontremoli", University of Milan, via Celoria 16, 20133 Milan, Italy}

\begin{abstract} Viscosity, a fundamental transport and rheological property of liquids, quantifies the resistance to relative motion between molecular layers and plays a critical role in understanding material behavior. Conventional methods, such as the Green-Kubo (GK) approach, rely on time integration of correlation functions, which becomes computationally intensive near the glass transition due to slow correlation decay. A recently proposed method based on non-affine lattice dynamics (NALD) and instantaneous normal mode analysis offers a promising alternative for estimating the viscosity. In this study, we apply the NALD approach to compute the viscosity of the Kremer-Grest polymer system over a range of temperatures and compare these results with those from the GK method and non-equilibrium molecular dynamics simulations. Our findings reveal that all vibration modes, including the instantaneous normal modes, contribute to the viscosity. This work presents an efficient framework for calculating viscosity across diverse systems, including near the glass transition where the GK method is no longer applicable. Also, it opens the avenue to understanding the role of different vibrational modes linked with structure, facilitating the design of materials with tunable rheological properties.\end{abstract}

\maketitle

\section{Introduction\label{Intro}}
The structure and dynamics of polymers are crucial for the development of different chemical and biological applications \cite{doi_1988}. Over the past decades, extensive studies on both naturally occurring and synthetic polymers, such as proteins and chemical compounds, have advanced our understanding in these areas. Polymer dynamics depends on the structural arrangement, length scales, and topological constraints imposed by the surrounding chains. Early work, such as single-chain studies by de Gennes \cite{DeGennes_1979}, laid the foundation to the understanding of these dynamics. Over the years, various groups have investigated the dynamical and rheological properties of polymer melts using theoretical, simulation, and experimental approaches, revealing phenomena such as contour length fluctuations \cite{Doi_1983, Milner_1998}, constraint release \cite{Rubinstein_1988, Shivokhin_2014}, collapse kinetics \cite{Byrne_1995, Kikuchi_2005, Chauhan_2022}, confinement effects \cite{McLeish_2002}, and phase behavior \cite{macrmolecule_2015, JCP_2021}. Recently, significant attention has been directed to the study of the rheological properties of entangled and ring polymer melts \cite{ACS_2013, ACS_2021, Parisi_2021, Datta_2023, roy2022effect, macmolecule_2024}. Polymer melts, with their amorphous properties, serve as effective models for supercooled liquids \cite{Dudowicz_2005, vaibhav2024entropic}. Although tube theory and reptation concepts have proven invaluable in obtaining a deeper molecular level understanding of polymer dynamics \cite{doi_1988}, for many applications and materials design it is becoming increasingly crucial to be able to quantitatively predict rheological properties from atomic-scale structure and interactions \cite{theodorou,Doros}, something that mean-field methods cannot achieve.

Viscosity is a fundamental property for understanding the transport behavior of liquids and supercooled liquids \cite{viswanath2007viscosity, Murillo,huang_plasma,Trachenko_2023, Zaccone_book,Tysoe}. In supercooled liquids, viscosity increases rapidly upon cooling, making it difficult to measure it near the glass transition temperature \cite{Ediger_1996, Debenedetti_2001}. Various methods are available to calculate the viscosity in liquids and amorphous systems, including equilibrium and non-equilibrium approaches \cite{Sen_2005,kroger_1993, Hess_2002, Evans_Morriss_2008}. In equilibrium methods, viscosity is derived from pressure or momentum fluctuations, while non-equilibrium methods rely on measuring the mechanical response to external deformation. The Green-Kubo (GK) method \cite{Green_1954, kubo_1957} is the most widely used equilibrium approach, whereby viscosity is calculated by integrating the stress autocorrelation function. These correlations decay slowly, and it is difficult to measure smooth long-time tails of such correlations. In general, non-equilibrium methods require very small shear rates to measure Newtonian viscosity, and also extensive averaging due to the highly fluctuating nature of the stress-strain behavior. Consequently, these methods are suitable within a limited temperature range for supercooled liquids. Furthermore, the theoretical approach to derive a microscopic expression for viscosity involves linking particle probability distributions and interatomic potentials using the Born-Green approximation \cite{Born_1947}. By employing a hole model of the liquid state, an analytical relationship between viscosity and the thermodynamic properties of dense liquids has been established through the determination of an equation of state \cite{Sanchez_1974, Locati_1999}. Other approaches, such as activation rate theory \cite{glasstone1941theory, Douglas_PANS_2009,Dyre,Schweizer_2005,Ankit_2021,Ankit_2023}, relate viscosity or relaxation times to the hopping rates of particles as they move across the cage formed by their nearest neighbors.

There have been attempts to relate the frequency-dependent shear modulus of viscoelastic liquids to viscosity. One such phenomenological framework is based on Maxwell’s model \cite{frenkl1955kinetic}, and several proposals have explored the connection between the viscoelastic modulus and transport properties, such as viscosity, in supercooled liquids \cite{Sen_2005, kroger_1993, Hess_2002, marrucci2000molecular, Peluso_2024, Huang_2025}. However, a significant theoretical advancement in this direction is the development of non-affine lattice dynamics (NALD), where the equation of motion for the non-affine displacement of a tagged particle in a disordered medium is explicitly solved, providing deeper insight into the microscopic mechanisms governing viscosity. The NALD approach relies on specific microscopic inputs, primarily the nature of the potential energy surface and the Hessian matrix of the system, to predict the frequency-dependent mechanical response of the system \cite{Zaccone_book}.

At its core, the NALD framework is based on the fundamental concept of non-affine displacements \cite{lemaitre_2006, Scossa}. When a material sample undergoes deformation, each atom tends to follow the applied strain, and the affine displacement is defined as the atomic displacement from its original position in the undeformed sample to the position prescribed by the macroscopic strain. In this (affine) position, due to the lack of centrosymmetry, the atom is not at mechanical equilibrium and must undergo an additional displacement, which is the non-affine displacement \cite{Slonczewski,theodorou}. The force acting on the atom in the affine position, which triggers the non-affine displacement, is called the affine force or affine force field \cite{Zaccone_book}. The solution, expressed in terms of the complex viscoelastic modulus as a function of external frequency, is then obtained after performing projection of the atomic displacements onto the eigenvectors of the Hessian matrix, and upon Fourier transformation. Finally, the viscosity is computed using the viscoelastic modulus within the framework of non-affine response theory \cite{Zaccone_2023}.

In this paper, we investigate the microscopic origin of viscosity in supercooled liquids from first principles using non-affine response theory based on the NALD framework. We study a coarse-grained polymer melt in the supercooled regime by employing molecular dynamics (MD) simulations to generate the configurations \cite{kremer1990dynamics, palyulin2018parameter}. These configurations are then used to calculate the vibrational density of states (vDOS) and affine force field correlators, which are subsequently used to compute the complex shear modulus. The viscosity of the system in the zero-frequency limit is directly obtained from the loss modulus following the theory of Ref. \cite{Zaccone_2023}. The viscosity in the supercooled regime and its temperature dependence is accurately captured by this NALD approach. We also calculate the viscosity using the GK method and non-equilibrium shear MD simulations over the accessible temperature range for supercooled liquids. By comparing the viscosity results obtained from different methods in both the liquid and supercooled regimes, we show that the NALD approach reliably predicts the viscosity, particularly in the supercooled regime where experimental measurements and simulations based on the conventional methods are challenging. 

In Section \ref{method}, we describe the model and discuss the simulation details of the polymer melts studied. We provide a brief overview of the NALD theory and a concise summary of the GK and NEMD methods employed in our analysis for comparison. In Section \ref{Result}, we present and analyze our results, highlighting their significance and comparing them with those obtained from different approaches. Finally, in Section \ref{Conclusion}, we summarize our findings, discuss their implications, and outline potential directions for future research.

\section{Model and methods \label{method}}
We have studied a coarse-grained polymer, popularly known as Kremer-Grest model \cite{grest1986molecular}, where the polymer chain has beads (monomers) connected linearly via non-breakable covalent bonds, represented by finite extensible nonlinear elastic (FENE) potential \cite{kremer1990dynamics}. Such a potential has the form $U_{\rm FENE} = -0.5 KR_{0}^{2} \ln [1 - (r/R_{0})^{2}]$. In this model, the monomers interact pairwise with all other monomers via a Lennard-Jones (LJ) interaction $U_{\rm LJ} = V(r) - V(r_c)$, where $V(r) = 4 \epsilon [(\sigma/r)^{12}-(\sigma/r)^{6}]$, for $r < \sigma$, and $r$ is the inter-monomer separation. All the measurements are performed in standard reduced LJ units. In this study, we set the parameters as follows: $K = 30, R_{0} = 1.5, \epsilon=1, \sigma=1, r_{c}=2.5$. The mass of each monomer is set to unity. We have only these two types of interaction present in the system. Since there are no bending potentials in the form of three monomer angle, four monomer dihedral, etc., the polymer chain is fully flexible. We have simulated $M = 100$ polymer chains, each consisting of $50$ identical beads. This makes a system of $N = 5000$ particles. 

Large-scale molecular dynamics simulations are performed using LAMMPS \cite{Plimpton.1995}, where equations of motion are integrated via velocity-Verlet algorithm with timestep $\Delta t = 0.005$ under periodic boundary conditions. To prepare an equilibrated configuration in the supercooled state, we start with a random configuration of a polymer chain, prepared via self-avoiding random walk. These random configurations are first heated at a sufficiently high temperature and zero pressure, and then quenched to the target temperature and equilibrated, maintaining the zero pressure. Langevin thermostat and Nosè-Hoover barostat, both having unit dissipation timescale, are employed to maintain the temperature and pressure of the system. For the system we have used in our study, the glass transition temperature is\cite{palyulin2018parameter, kremer1990dynamics} $T_g \simeq 0.4$. We have generated at least $30$ independent configurations at each temperature in the range $\left[0.6,1.8\right]$.

In this work, we have also studied the response of the system to shear deformation. Shear is imposed by deforming the $xy$-plane in the $x$-direction \cite{vaibhav2022rheological,vaibhav2023controlled}, at different shear rates $\dot{\gamma}$ ranging between $10^{-4}$ and $5 \times 10^{-3}$. In the presence of shear, Lees-Edwards boundary conditions are utilized to take into account the deformed boundary \cite{allen2017computer}. Furthermore, during deformation, the temperature of the system is maintained using dissipative particle dynamics thermostat \cite{soddemann2003dissipative, vaibhav2021influence} without any external pressure control. Shear stress $\sigma_{xy}$ is measured using the virial definition: $\sigma_{xy} = \langle \frac{1}{V} \sum_{i, j} f^{x}_{i,j} r_{i,j}^y \rangle$, where $f^{x}_{i,j}$ is the $x$-component of the force acting on particle $i$ due to $j$, $r_{i,j}^y$ is the $y$-component of the inter-particle separation between $i$ and $j$, and $V$ is the volume of the system.\cite{thompson2009} We study the evolution of shear-stress $\sigma_{xy}$ as a function of deformation in the form of shear-strain $\gamma$ at different $\dot{\gamma}$.

For the normal mode analysis, we compute the dynamical matrix or Hessian ${\bf H}$, using the method of finite-differences. Instantaneous configurations at different temperatures are utilized for these calculations. For our three-dimensional system of $N$ particles, the Hessian is a $3N \times 3N$ matrix. The elements of the matrix are defined as ${\bf H}_{ij}^{\alpha \beta} = \frac{1}{\sqrt{m_im_j}} \frac{\delta^2 U}{\delta r_{i,\alpha}\delta r_{j,\beta}}$, where $i,j$ are particle index, $\alpha, \beta$ are Cartesian directions, and $U$ is the system potential energy. We diagonalize the matrix using the LAPACK package implemented in Intel MKL library \cite{anderson1999lapack}. This results in $3N$ eigenvalues $\lambda_{l}$ and the same number of eigenvectors ${\bf e}_{l}$, $l = {1, 2, 3, ..., 3N}$. An eigenfrequency $\omega$ is defined corresponding to each eigenvalue, $\omega_l = \sqrt{\lambda_{l}}$. Since we study finite temperature configurations, negative eigenvalues are present after diagonalization, which result in imaginary eigenfrequencies \cite{palyulin2018parameter, vaibhav2024time}, also known as instantaneous normal modes (INMs) \cite{keyes1994unstable,stratt1995instantaneous,Laird,Jack}.

\subsection{Non-affine lattice dynamics approach to viscosity} \label{NALD}
In an ideal elastic solid, when deformation is applied, particles move toward their affine positions and the local inversion symmetry ensures that forces from neighboring particles cancel out, leaving the particle in mechanical equilibrium at the affine position. However, the basis of non-affine lattice dynamics (NALD) lies in recognizing that, in disordered (amorphous) solids, the standard affine approximation of lattice dynamics breaks down. In such systems due to the absence of centrosymmetry, the applied deformation generates non-zero net forces on atoms, leading to additional non-affine displacements. This process requires internal work to displace particles from their affine to non-affine positions, which contributes negatively to the free energy of deformation. The equation of motion for a tagged particle $i$ of mass $m$ within the framework of the generalized Langevin equation \cite{cui_2017,palyulin2018parameter, zwanzig2001nonequilibrium} reads as
\begin{align}
    m \Ddot{\textbf{r}}_{i}+ \nu \dot{\textbf{r}}_{i} + \sum_j \textbf{H}_{ij} \textbf{r}_{j} = \Xi_{i} \gamma.
    \label{eq1}
\end{align}
Here, $\textbf{r}_{i}$ represents the position of the $i$th particle, $\nu$ is the friction coefficient that characterizes energy dissipation due to viscous damping at the atomic scale\cite{zwanzig2001nonequilibrium}, the third term (with ${\bf H}_{ij}$ the Hessian matrix) represents the sum of the  restoring forces arising from interactions with neighboring particles, and the term containing $\Xi$ represents the affine force induced by the applied shear strain $\gamma$.

The solution of Eq.~(\ref{eq1}) is obtained by performing a Fourier transform over normal mode decomposition, which eventually describes the complex viscoelastic modulus dependent on deformation frequency $(\Omega)$: 
\begin{align}
    G^{*}(\Omega) = G_{A}-\frac{1}{V}\sum_k \frac{\Gamma(\omega_{k})}{m(\omega^{2}_{k}-\Omega^{2})+i\Omega{\nu}},
    \label{G}
\end{align}
where, $G_{A}$ is the modulus due to affine part, $V$ is volume of system,  $\omega_{k}$ are the normal modes of system and $\Gamma(\omega_{k})=\Tilde{\Xi}^{2}_{k}$ is the affine force field correlator \cite{lemaitre_2006, Milkus_2018, palyulin2018parameter}. Equation \ref{G} can be expressed in a continuous form by using the normalized vibrational density of states $g(\omega)$
\begin{align}
    G^{*}(\Omega) = G_{A}-\frac{N}{V}\int \frac{\Gamma(\omega) g(\omega)}{m(\omega^{2}-\Omega^{2})+i\Omega{\nu}} d\omega.
    \label{Gint}
\end{align}

By separating the real and imaginary parts of Eq.~(\ref{Gint}), we obtain the storage modulus ($G^{\prime}$) and the loss modulus ($G^{\prime \prime}$) \cite{Zaccone_2023,palyulin2018parameter},
\begin{align}
    G^{\prime} = G_{A}-\frac{N}{V}\int \frac{m \Gamma(\omega) g(\omega) (\omega^{2}-\Omega^{2})}{m^{2}(\omega^{2}-\Omega^{2})^{2}+\Omega^{2}{\nu}^{2}} d\omega,
    \label{Gp}
\end{align}
\begin{align}
    G^{\prime \prime} = \frac{N}{V}\int \frac{\Gamma(\omega) g(\omega) {\nu} \Omega} {m^{2}(\omega^{2}-\Omega^{2})^{2}+\Omega^{2}{\nu}^{2}} d\omega.
    \label{Gdp}
\end{align}
The vibrational density of states (vDOS), $g(\omega)$, is defined using the normal modes obtained after the diagonalization of the Hessian matrix constructed with the instantaneous configuration of the system \cite{Zaccone_book, Hara_2025}: $g(\omega)=\frac{1}{3N-3}\sum_{i}\delta(\omega-\omega_{i})$, where we discard three Goldstone modes appearing due to rigid-body translations. Such vDOS can be derived analytically \cite{zaccone_PNAS_2021} or computed numerically \cite{Jack,palyulin2018parameter}, and can also be extracted from experimental data \cite{Stamper_2022,vaibhav2025experimental}. 
The frequency-dependent affine force field correlator, $\Gamma(\omega) $, is calculated as the square of the projection of the force field $\Xi$ onto the eigenvectors corresponding to the normal modes. The affine force field, which drives non-affine motion in the sheared system, is calculated as the net force acting on each atom after the system is affinely deformed by a very small shear strain $\eta=10^{-6}$. Specifically, \(\Xi_{i} = \Delta f_{i} / \eta\), where $\Delta f_i$ represents the net force on atom $i$ relative to its position in the undeformed state. The analytical estimate of $\Gamma(\omega)$ at low eigenfrequencies, derived in Ref. \cite{Scossa}, is $\Gamma(\omega)  \sim \omega^{2}$, and has been proved to be a reasonable approximation for amorphous solids \cite{Pastewka,flenner2024}. To calculate the affine modulus \( G_A \), the system is deformed affinely with a small shear strain \( \eta = 10^{-6} \) and the corresponding shear stress \( \sigma \) is measured. Then, the affine modulus is calculated as $G_A = \sigma/\eta$.

The non-affine response theory, developed from first principles, establishes a relationship between the shear viscosity and the loss modulus $G^{\prime\prime}$, given by \cite{Zaccone_2023}
\begin{align}\nonumber
    \eta =& \frac{G^{\prime\prime}(\Omega)}{\Omega}|_{\Omega\rightarrow 0},\\
    =& \frac{N {\nu}(0)}{V}\int \frac{\Gamma(\omega) g(\omega)} {m^{2}\omega^{4}} d\omega.
    \label{viscosity_eq}
\end{align}
The relation provides a direct link between the viscosity $\eta$, the vDOS $g(\omega)$, and the affine force field $\Gamma(\omega)$, all of which are computed numerically for our system.

Throughout this study, we use a fixed value of ${\nu} = 1$ in the above equations, which matches exactly the value implemented in the Langevin thermostat of the MD simulations \cite{palyulin2018parameter}. Hence, there are no adjustable parameters in our calculations. One could consider varying the memory friction parameter $\nu$ and observing its effect on the viscoelastic modulus \cite{palyulin2018parameter} and thermodynamic properties. In a Markovian process, friction is treated as a constant, as implemented in molecular simulations using the Langevin thermostat. However, in real materials such as metallic glasses or experimental systems, friction is better described within a non-Markovian framework, where the friction kernel is time-dependent or history-dependent \cite{Klippenstein_2021, Milster_2024}. The friction kernel can be computed using the fluctuation-dissipation theorem (FDT), allowing us to determine the memory kernel in the frequency domain. This frequency-dependent memory kernel can then be incorporated into the viscoelastic modulus, providing a more accurate description of real materials and experimental conditions, which is a whole task in its own right that we leave for future studies.

\subsection{Green-Kubo method for viscosity calculation}
The Green-Kubo (GK) method \cite{Green_1954,kubo_1957} is the most widely used technique to calculate transport coefficients, such as heat conductivity \cite{vaibhav2020response}, shear viscosity, and bulk viscosity, relating the time correlation functions to the corresponding fluxes or tensors in thermal equilibrium. To calculate the shear viscosity using this method, at first the autocorrelation function of stress $\sigma_{\alpha\beta}$ (with $\alpha, \beta = x, y, z$) is calculated as a function of time with the equilibrium trajectory of the system. The viscosity is then obtained by integrating the stress autocorrelation function over time, and the expression is given by \cite{Helfand_1960}
\begin{align}\label{GK}
    \eta = \frac{V}{k_{B}T}\int_{0}^{\infty} \langle \sigma_{\alpha\beta}(t)\sigma_{\alpha\beta}(0)\rangle dt.
\end{align}
Here, $T$ and $V$ are the temperature and volume of the system, $k_B$ is the Boltzmann constant, and the brackets $\langle \cdot \rangle$ denote the ensemble average. Due to the isotropy of the system, the three independent off-diagonal components of the stress tensor are expected to be equivalent; hence, the final viscosity is obtained by averaging the three independent calculations based on the autocorrelation of these off-diagonal stress components. Moreover, to further reduce the noise level in the autocorrelation function we also perform running time average \cite{Sen_2005}. Despite this simplification, the accuracy of the GK method in MD simulations is limited, especially at small temperatures, because at these temperatures the autocorrelation function decays very slowly and the long-time tail of the correlation function is not sampled very well. 

\subsection{Viscosity via non-equilibrium molecular dynamics}
Another very popular method to estimate viscosity is via non-equilibrium molecular dynamics (NEMD) \cite{thin}. In this approach, the system is subjected to shear deformation at a rate $\dot{\gamma}$, and the resulting shear stress is measured as the deformation progresses. Initially, the response is linear, followed by a nonlinear regime, until the system reaches a steady state where the shear stress fluctuates around a mean value. If the system is sheared along the $xy$-plane, the stress response is given by $\sigma_{xy}$,  and the steady-state average shear stress for a given shear rate is denoted as $\langle \sigma_{xy} \rangle_{\rm S}$. The shear viscosity $\eta$ is then defined as: $\eta = \langle \sigma_{xy} \rangle_{\rm S}/\dot{\gamma}$.

When viscosity is plotted as a function of shear rate, its behavior depends on the nature of the fluid. If viscosity decreases with increasing shear rate, the material exhibits shear-thinning behavior \cite{Hara_2025,thin}, whereas an increase in viscosity with shear rate indicates shear-thickening behavior \cite{morris2020shear}. A constant viscosity across shear rates characterizes Newtonian behavior. For most of the complex fluids, shear-thinning/shear-thickening vanishes at sufficiently small shear-rates and viscosity becomes a constant independent of shear-rate. Such a viscosity is called zero-shear viscosity given by,

\begin{equation}\label{Shear_MD}
    \eta = \lim_{\dot{\gamma} \to 0} \langle \sigma_{xy} \rangle_{\rm S}/\dot{\gamma}.
\end{equation}

\section{Results and Discussion}\label{Result}
\begin{figure}[t!]
    \centering
    \includegraphics[width=0.9\linewidth,clip]{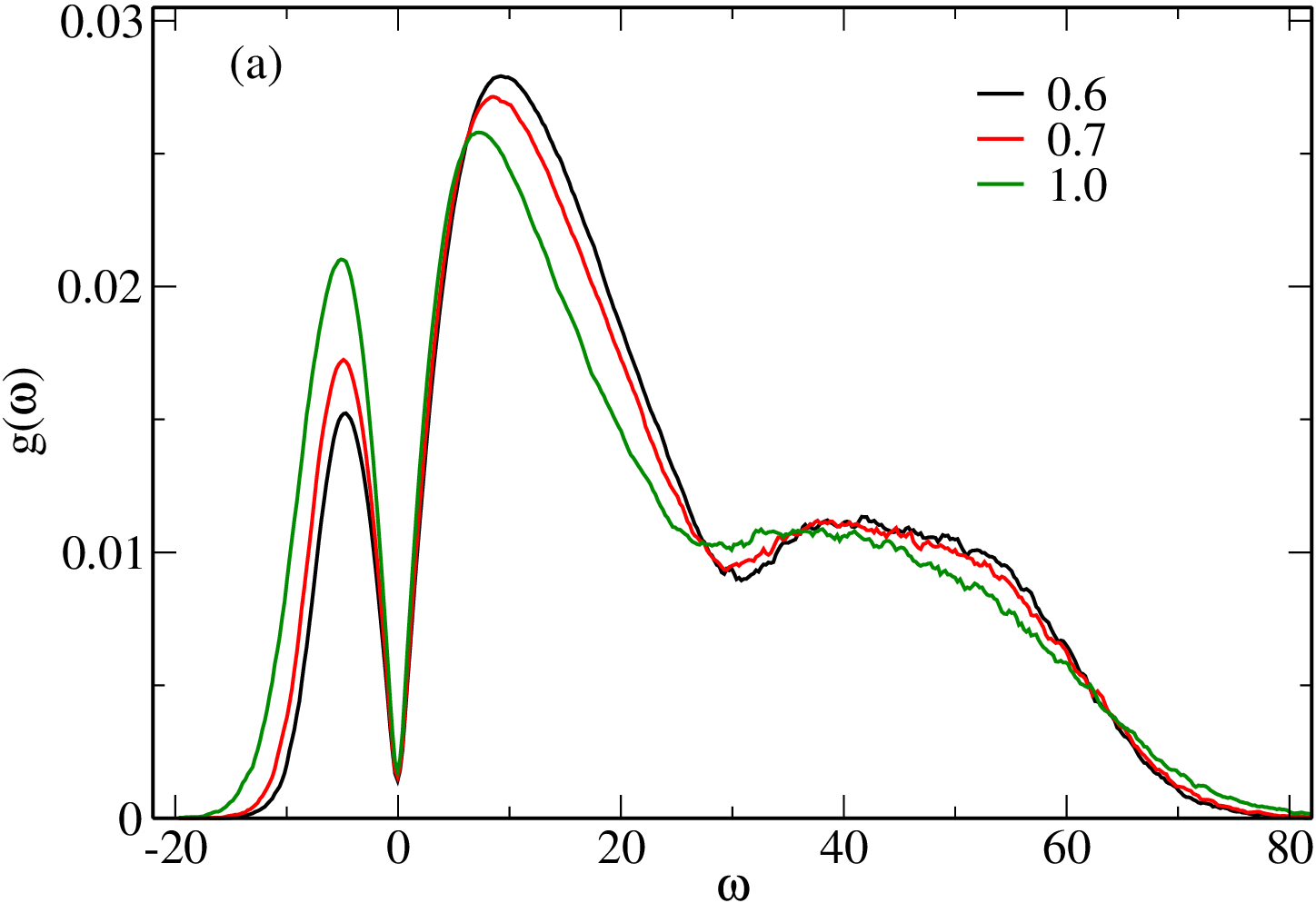}
    \centering
    \includegraphics[width=0.9\linewidth,clip]{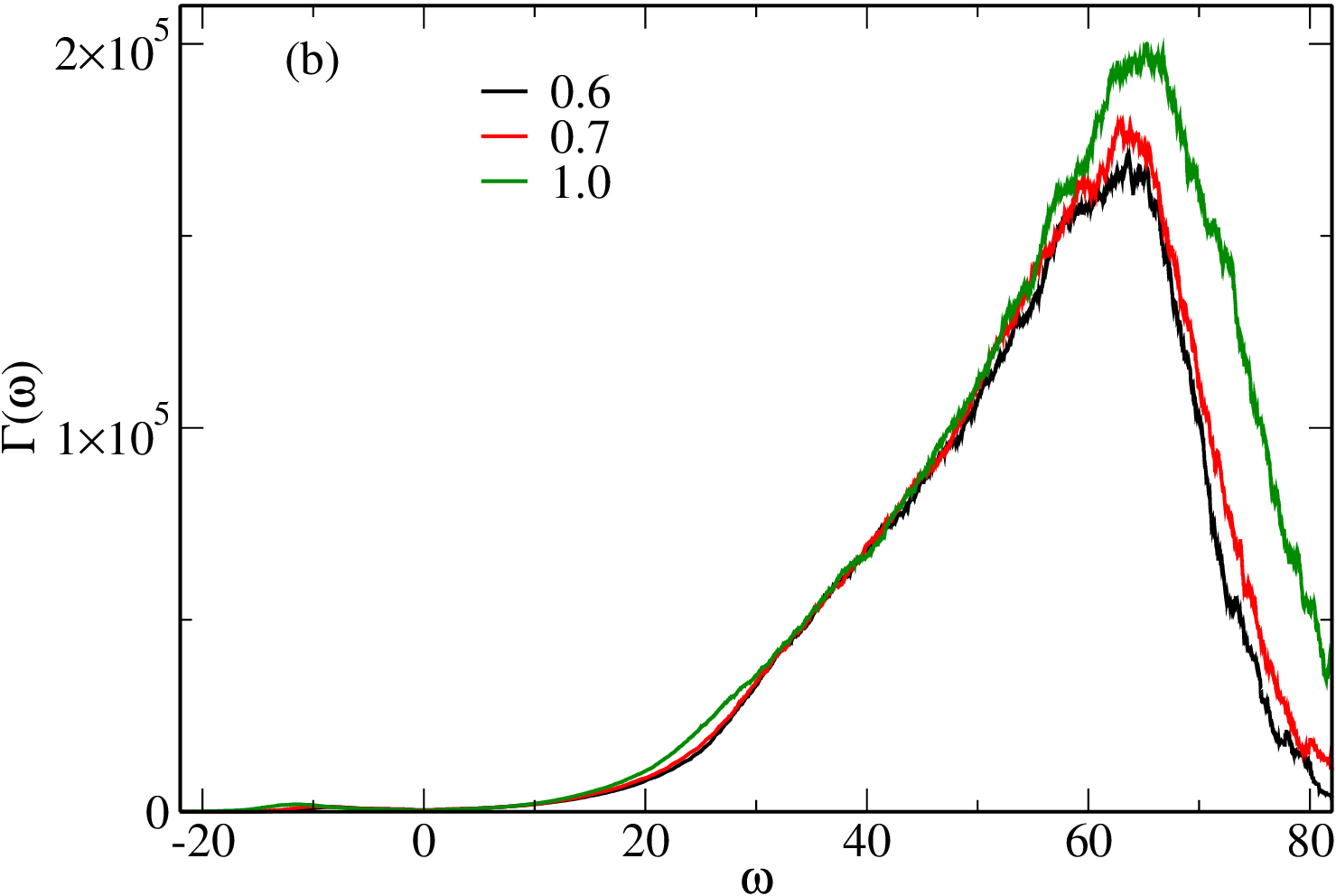}
    \caption{(a) Vibrational density of states $g(\omega)$ and (b) affine force field correlator $\Gamma(\omega)$ as a function of normal mode frequency $(\omega)$ at different temperatures (marked). The numerical calculation is done for the KG polymer system described in the text. The quantities are in Lennard-Jones units.}
    \label{DOS}
\end{figure}

We applied the framework discussed in Section \ref{NALD}, utilizing the microscopic formula Eq. \eqref{viscosity_eq} to establish the relationship between viscosity and normal mode analysis. The vibrational density of states is plotted in Fig.~\ref{DOS}(a) for three different temperatures. Since our system is at a finite temperature, the vDOS has a significant number of imaginary modes (conventionally shown on the negative branch of $\omega$) along with the real modes\cite{keyes1994unstable,stratt1995instantaneous,Jack}. The imaginary modes appear due to the presence of saddles in the energy landscape. As temperature decreases, the number of imaginary modes also decreases, whereas the number of real modes increases. There are two peaks on the positive side of the frequency axis, largest peak occurring at a smaller frequency (close to $\omega = 10$) is attributed to the vibration arising from the LJ interactions between beads, while the smaller peak is slightly broader and occurs at higher frequency (close to $\omega = 50$), primarily attributed to the vibrations corresponding to non-breakable FENE bonds  \cite{Pablo_2004, Milkus_2018}. As expected, \( g(\omega) \) exhibits a linear behavior at low frequencies, consistent with both theoretical predictions \cite{zaccone_PNAS_2021} and experimental observations \cite{Stamper_2022, vaibhav2025experimental}.

In Fig.~\ref{DOS}(b), the affine force field correlator \( \Gamma(\omega) \) is plotted as a function of the normal mode frequency \( \omega \) for three different temperatures. We observe that \( \Gamma(\omega) \) starts increasing rapidly for \( \omega > 20 \) and exhibits a temperature-dependent peak near \( \omega \simeq 70 \). The peak intensity decreases as the temperature decreases and shifts towards lower frequencies. The decay of the correlator function is also temperature-dependent; faster decay is observed at smaller temperatures. At low frequencies, \( \Gamma(\omega) \) is expected to follow the analytical behavior\cite{Scossa,Milkus_2017} \( \Gamma \sim \omega^2 \).

\begin{figure}[t!]
    \includegraphics[width=0.9\linewidth,clip]{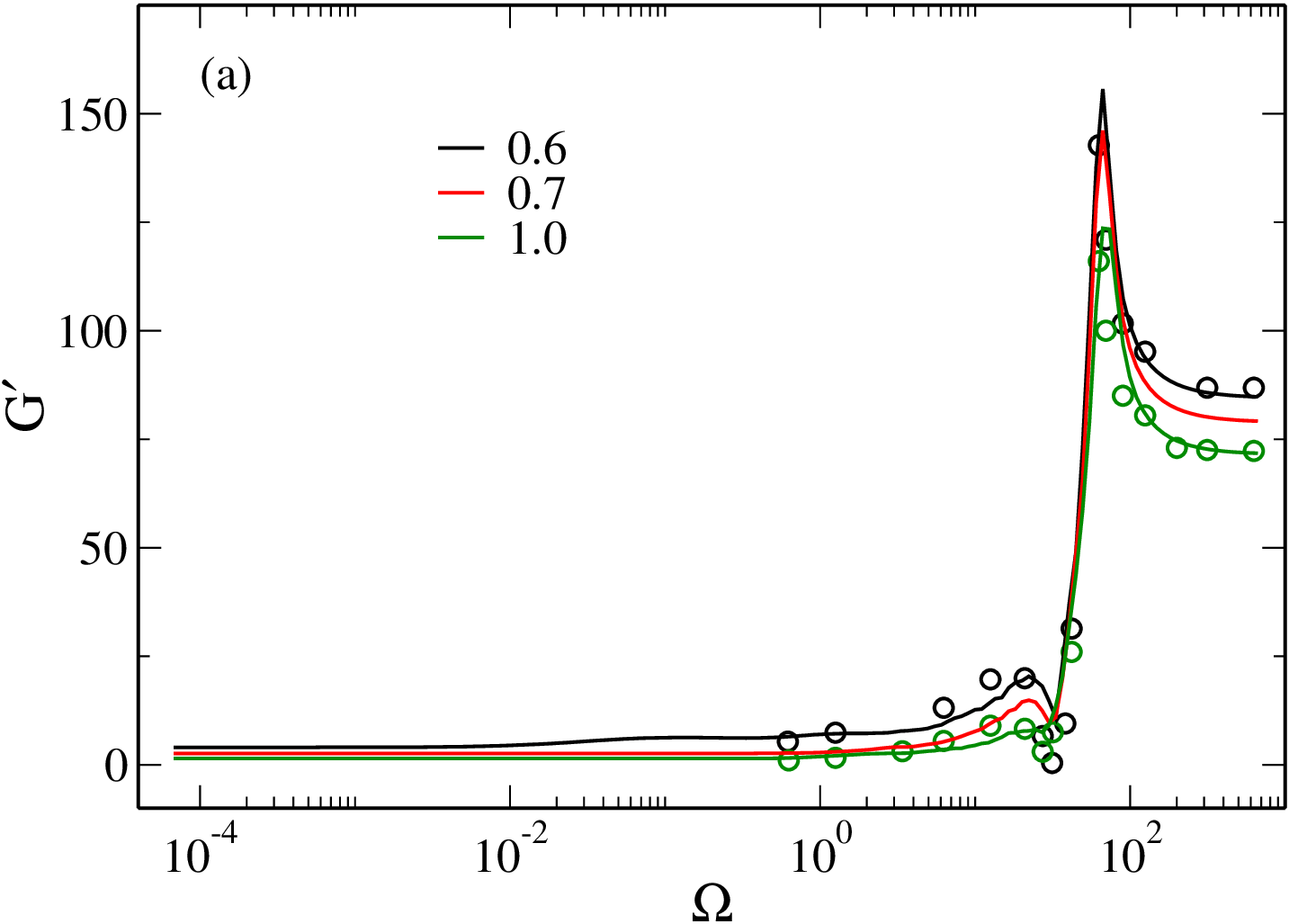}
    \includegraphics[width=0.9\linewidth,clip]{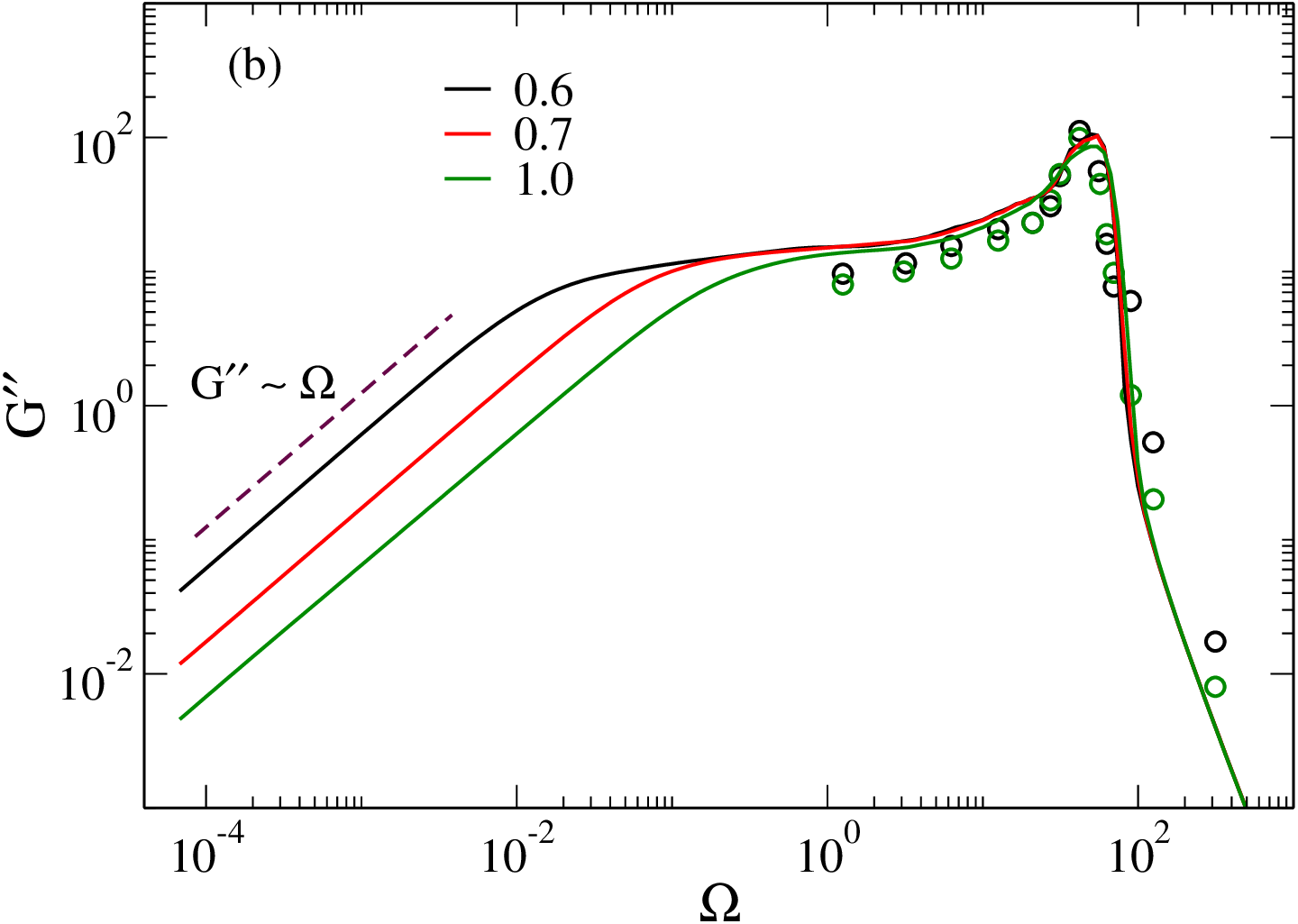}
    \caption{Dependence of the (a) real ($G^{\prime}$, storage modulus) and (b) imaginary ($G^{\prime \prime}$, loss modulus) parts of the complex viscoelastic modulus $(G^{*})$ on external deformation frequency $\Omega$ at different temperatures (marked). Lines represent NALD results and symbols represent MD results obtained from oscillatory shear. The color of the symbols matches that of the lines for the same temperature (only $T=0.6$ and $1.0$ are reported for the MD). The MD simulations are in parameter-free agreement with NALD predictions up to the accessible frequency range. The quantities are in Lennard-Jones units.}
    \label{G_dprime}
\end{figure}

\begin{figure}[t]
    \centering
    \includegraphics[width=0.9\linewidth,clip]{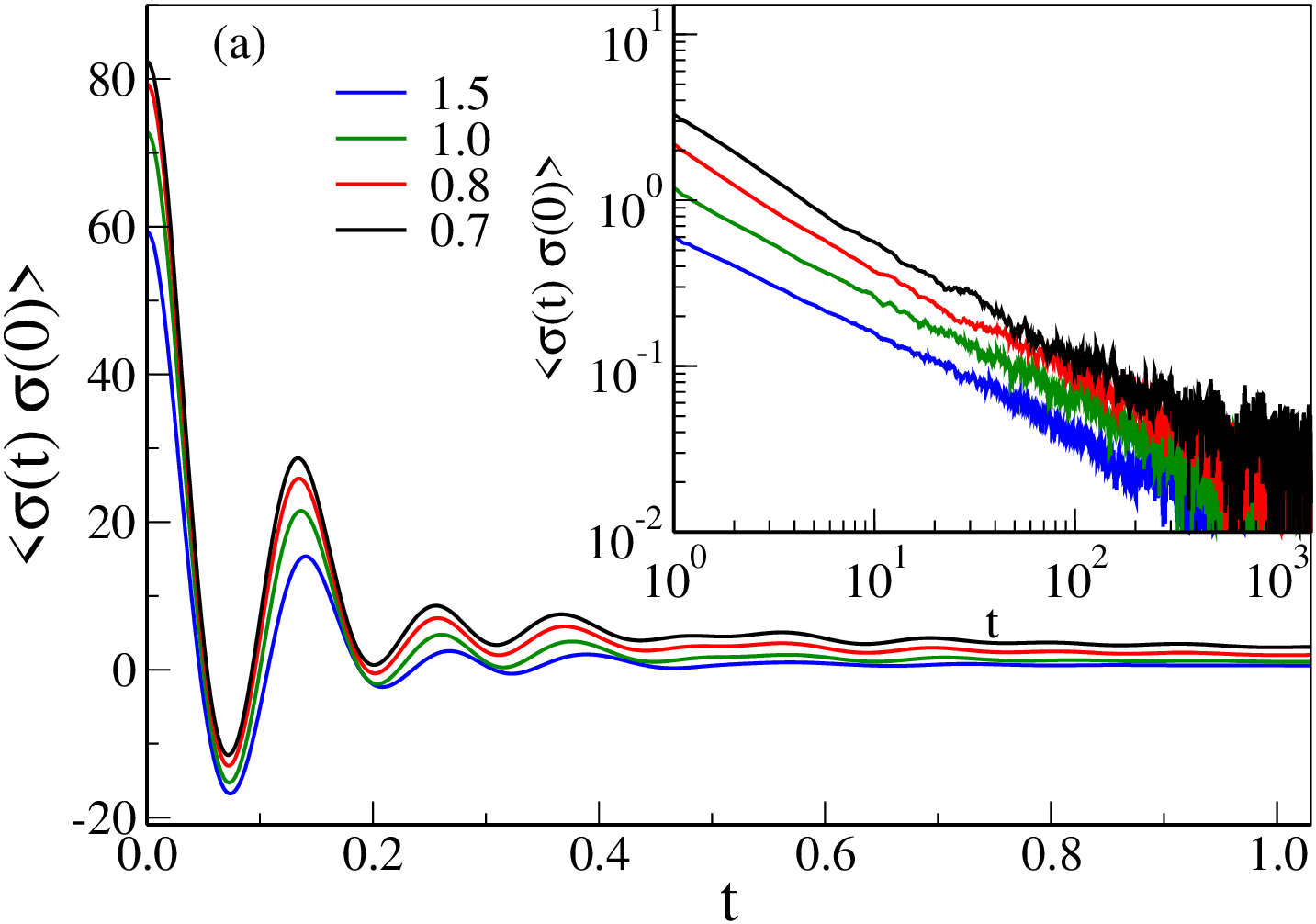}
    \includegraphics[width=0.9\linewidth,clip]{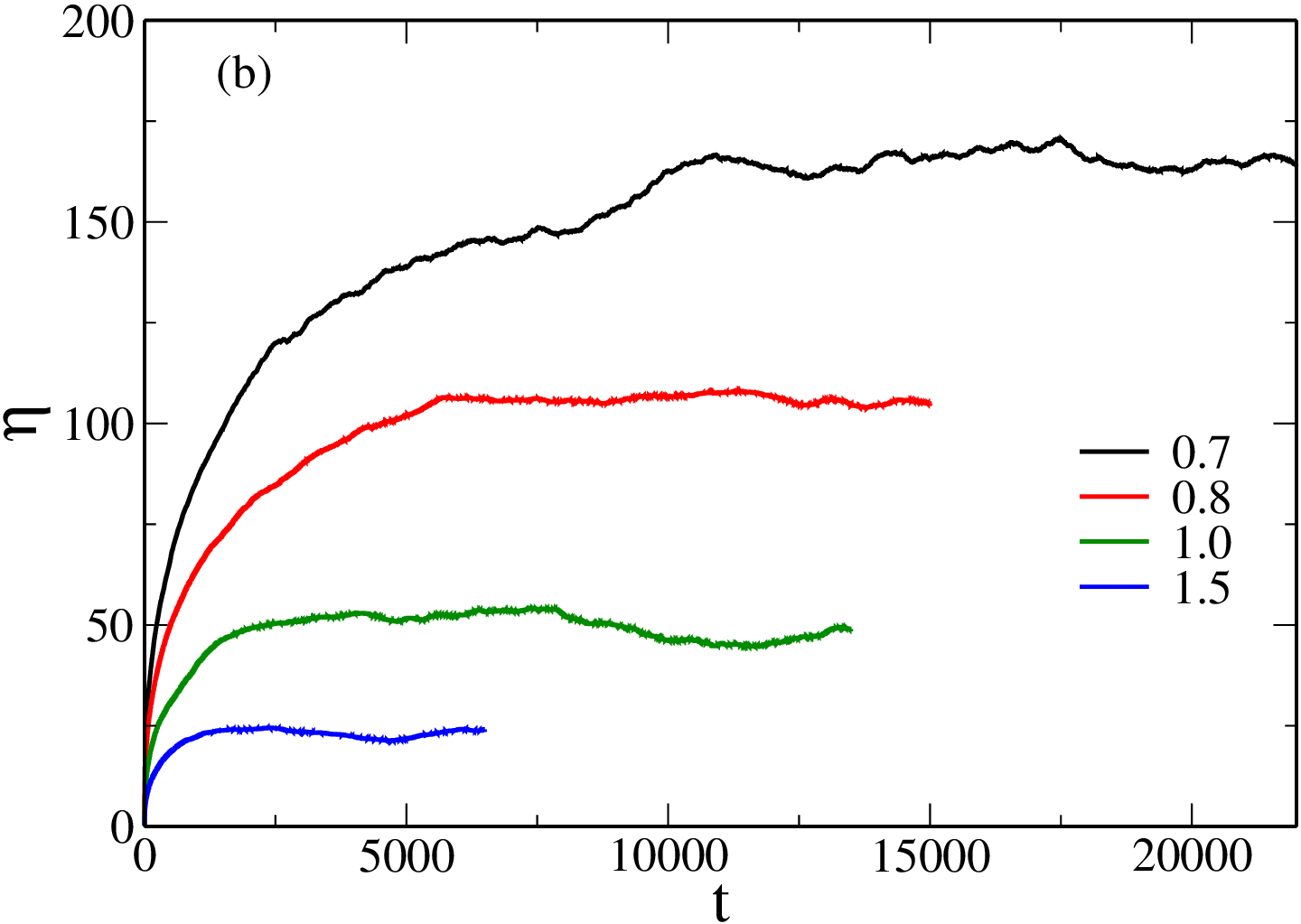}
    \caption{(a) Stress autocorrelation function at different temperatures (marked). Inset shows the tail part of the correlation function. (b) Plot of Green-Kubo viscosity as a function of time at different temperatures (marked). The quantities are in Lennard-Jones units.}
    \label{viscosity_with_time}
\end{figure}

Using the information of $g(\omega)$ and \( \Gamma(\omega) \) we calculate the viscoelastic modulus as a function of external frequency $\Omega$ in the liquid and supercooled regions. In Fig.~\ref{G_dprime} we plot real and imaginary parts, $G^{\prime}$ and $G^{\prime \prime}$, which represent the storage and loss moduli respectively, as a function of external frequency $\Omega$ at different temperatures.  $G^{\prime}$ exhibits specific features: there exists two different resonating frequencies where the response is significantly high with respect to the response at nearby frequencies. The first resonating peak, at a lower frequency, occurs at a value of $\Omega$ that matches with the vibrational eigenfrequency of the LJ interaction. Below this resonance frequency, $G^{\prime}$ decreases monotonically upon decreasing the frequency $\Omega$. The second resonance peak is higher than the previous peak and occurs at a larger value of $\Omega$, close to the characteristic vibration corresponding to FENE bonds\cite{palyulin2018parameter,Milkus_2018}. Since these bonds are non-breakable and more energetic, the mechanical response at $\Omega$ corresponding to FENE bonds is strongest. At the higher-frequency side, greater than any resonating frequencies, there exists a temperature-dependent plateau, representing the high-frequency modulus dominated by purely affine displacements.


In Fig.~\ref{G_dprime}(b), we plot $G^{\prime \prime}$ as a function of $\Omega$ at three temperatures. We observe that the value of $G^{\prime \prime}$ becomes very small at both low and large frequencies. It exhibits a broad peak at $\Omega \sim 70$, and below this frequency $G^{\prime \prime}$ decreases, forming a plateau having a wider width at smaller temperatures due to slower decay in the low-frequency regime. Subsequently, $G^{\prime \prime}$ decreases linearly with $\Omega$ as $G^{\prime \prime} \sim \Omega$ and continues as $\Omega$ approaches the zero-frequency limit \cite{Hara_2025}. 

In the evaluation of viscosity using the NALD approach (Eq. \ref{viscosity_eq}), we perform the integration over both real and imaginary branches of $\omega$. As shown in Ref.\cite{palyulin2018parameter}, it is important to include the imaginary modes in the calculation to capture the effect of temperature. A fraction of small-frequency modes below a cutoff frequency $\omega_{\rm min}$ is discarded during the integration, because these modes are unphysical due to the finite system size, as the system cannot support propagating modes corresponding to frequencies below the speed of sound in the medium. Such a minimum frequency can be estimated as $\omega_{\rm min}={2 \pi}/{L}\sqrt{{G_{s}}/{\rho}}$, where $L$ is the size of the simulation box, $\rho$ is the mass density, $G_{s}$ is the zero-frequency shear modulus \cite{vaibhav2024time}. The zero-frequency shear modulus is calculated by subtracting the shear stress fluctuation from the infinite-frequency shear modulus ($G_{\infty}$), i.e., $G_{s} = G_{\infty} - (V/k_{B}T) \left(\langle \sigma_{xy}^{2}\rangle-\langle\sigma_{xy}\rangle^{2}\right)$, cf. Ref.\cite{Ilg_2015}. The infinite-frequency shear modulus is calculated by deforming the system by a small shear-strain in the linear response regime, and the associated shear-stress is measured including both kinetic and virial components.

Another way to calculate the complex modulus \( G^* \) is through non-equilibrium molecular dynamics simulations under oscillatory shear \cite{palyulin2018parameter,vaibhav2024time,sirk2016bi}. In this method, a small-amplitude oscillatory shear strain \( \gamma(t) = \gamma_0 \sin(\Omega t) \) is applied to the system, and the corresponding stress response \( \sigma(t) = \sigma_0 \sin(\Omega t + \delta) \) is measured, where \( \Omega \) is the deformation frequency, and \( \delta \) is the phase difference between strain and stress. The \( G' \) and \( G'' \), which constitute the complex modulus \( G^* \), are calculated using the relations $G' = \left( {\sigma_0}/{\gamma_0} \right) \cos \delta, \quad G'' = \left( {\sigma_0}/{\gamma_0} \right) \sin \delta$. In these simulations, the strain amplitude \( \gamma_0 \) is kept very small and constant, and a thermostat is applied to maintain a constant temperature during deformation. It has been observed that the data for \( G' \) and \( G'' \) obtained from the small amplitude oscillatory shear (SAOS) simulations are consistent with the NALD approach at sufficiently higher frequencies. Within the available computational resources, generally, it is not possible to probe the response via SAOS at smaller frequencies accessible in experiments \cite{vaibhav2024time}. Therefore, the estimation of viscosity via SAOS, which requires measurement of \( G'' \) at small frequencies, becomes not viable.

We validate the NALD predictions of the viscoelastic modulus against non-equilibrium molecular dynamics (MD) simulations of oscillatory shear, where the \(xy\)-plane of the simulation box is sheared with an amplitude \(\gamma_0 = 0.01\) and frequency \(\Omega\) while measuring the stress response \(\sigma_{xy}\) as a function of time at temperatures \(T = 1.0, 0.6\).  For a reliable estimate, we average the stress response over \(10\) to \(30\) independent trajectory runs, after excluding the first \(30\) cycles of shear deformation. We use stress data and fit trigonometric functions to extract \(\sigma_0\) and the phase shift \(\delta\), allowing us to compute \(G'\) and \(G''\).  In Figs.~ \ref{G_dprime}(a) and \ref{G_dprime}(b), we compare the storage and loss moduli, as a function of $\Omega$ at temperatures \(T = 1.0\) and \(0.6\), obtained from the NALD approach (lines) with the MD results (symbols). The comparison shows a consistent parameter-free match between NALD and MD down to the accessible frequency range. No adjustable parameter is used in the comparison, because the value of $\nu$ used in the NALD equations and in the Langevin thermostat of the MD simulations is the same.

The calculation of zero shear rate viscosity using the GK formula involves the stress autocorrelation function (ACF) as a function of time (see Eq.~\ref{GK}). We show the temporal dependence of stress ACF in Fig.~\ref{viscosity_with_time}(a) at different temperatures. The initial decay of ACF is sharp and then it shows short-time oscillatory behavior, which is primarily attributed to the rapid fluctuations of the stiff covalent bond potential. In the inset of Fig.~\ref{viscosity_with_time}(a), we show the long-time tails of the ACF, which decays slowly and is highly fluctuating in nature. To reduce the noise in the long-time tails, we averaged the measurements over several time origins and independent replicas. Despite this precaution, it is not possible to significantly reduce the noise in the tails of ACF, leading to very slow convergence of integral in Eq.~\ref{GK} and hence making the computation costly.

\begin{figure}[t]
    \centering
    \includegraphics[width=0.9\linewidth,clip]{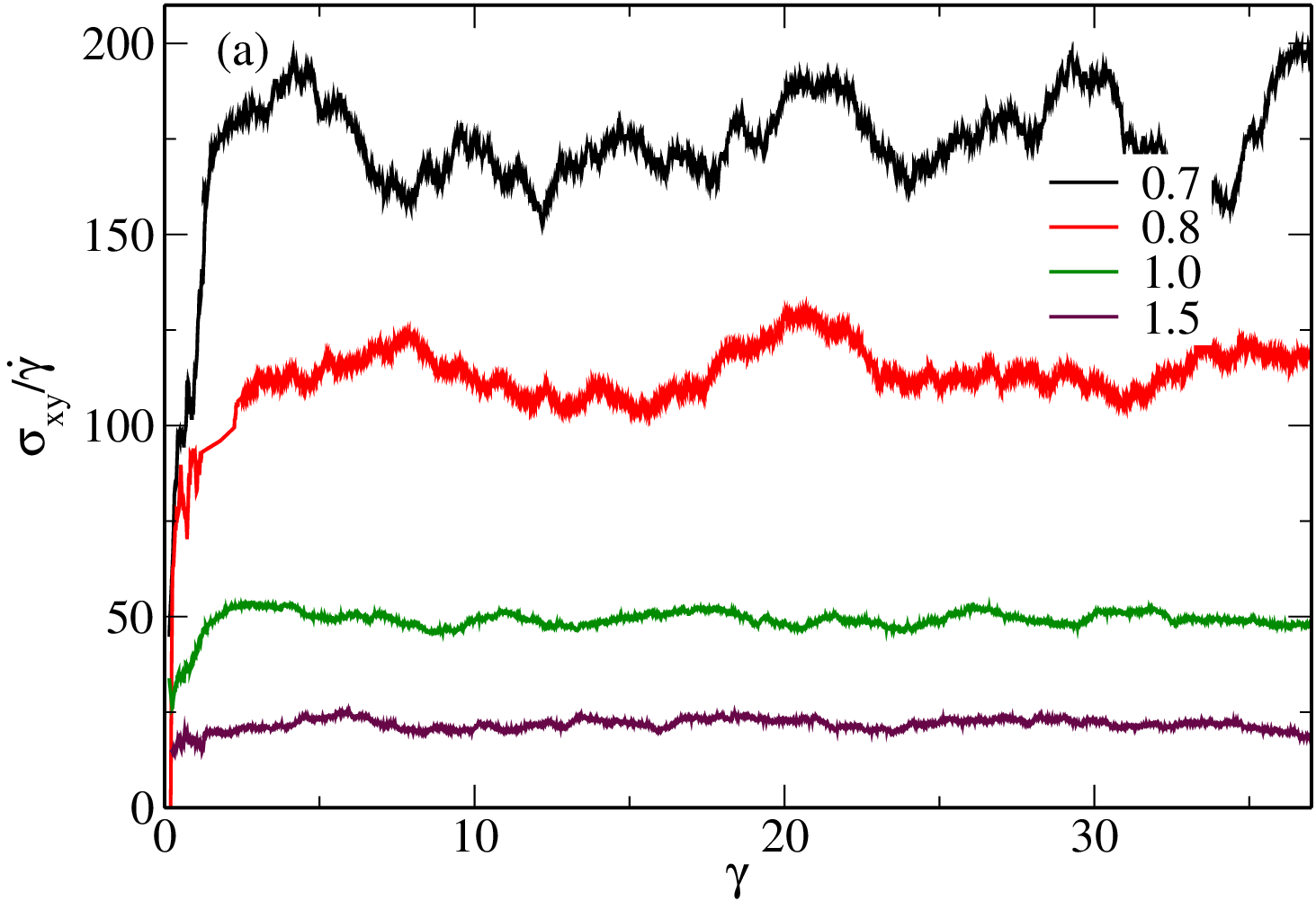}
    \includegraphics[width=0.88\linewidth,clip]{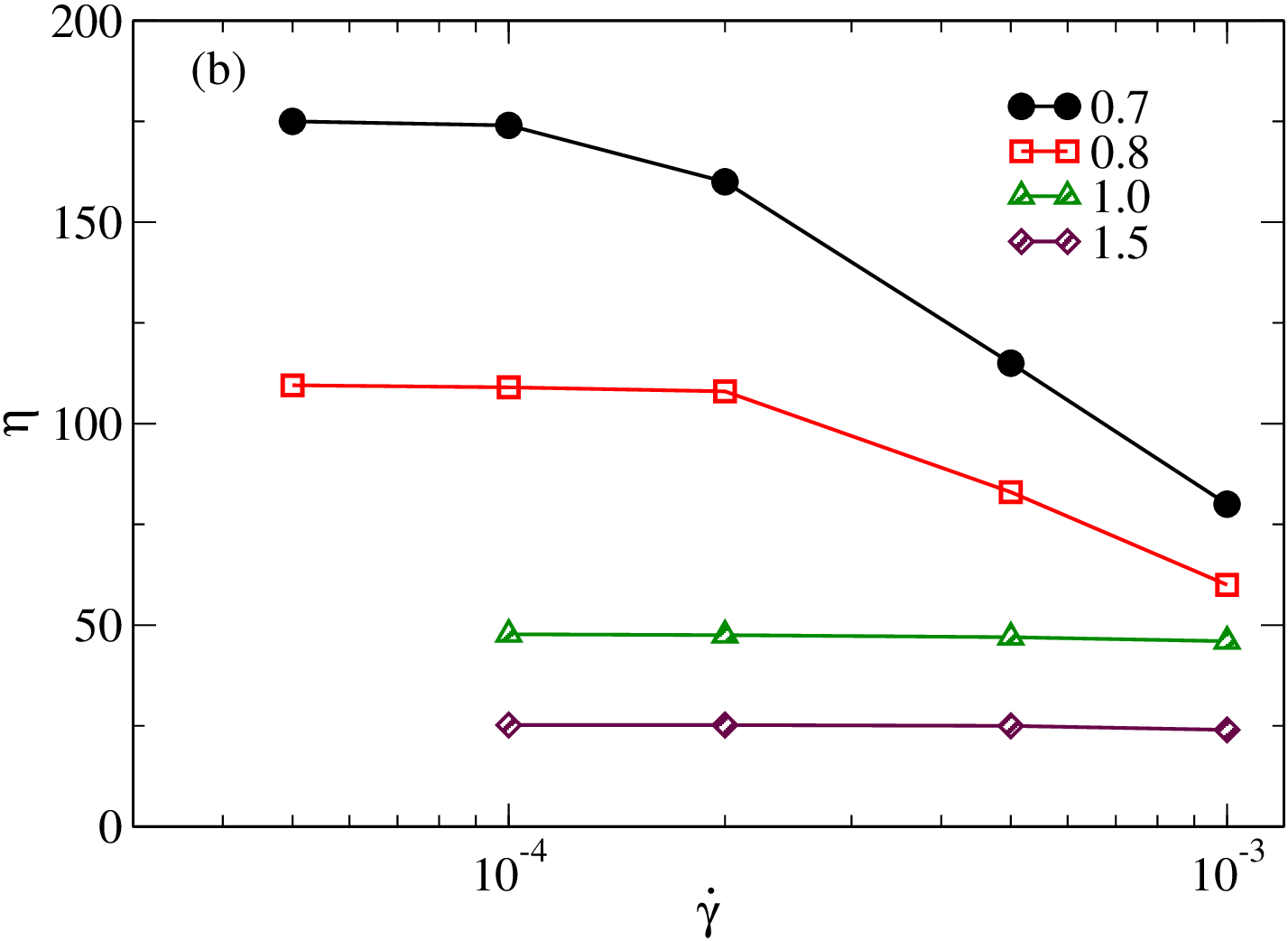}
    \caption{(a) Plot of $\sigma_{xy}/\dot{\gamma}$ i.e., the ratio of shear-stress to shear-strain rate, as a function of shear-strain $\gamma$ for different temperatures (marked). (b) Plot of shear viscosity $\eta$ as a function of shear rate $\dot{\gamma}$ at different temperatures (marked). The quantities are in Lennard-Jones units.}
    \label{stress-strain}
\end{figure}

\begin{figure}[t]
    \centering
    \includegraphics[width=0.98\linewidth,clip]{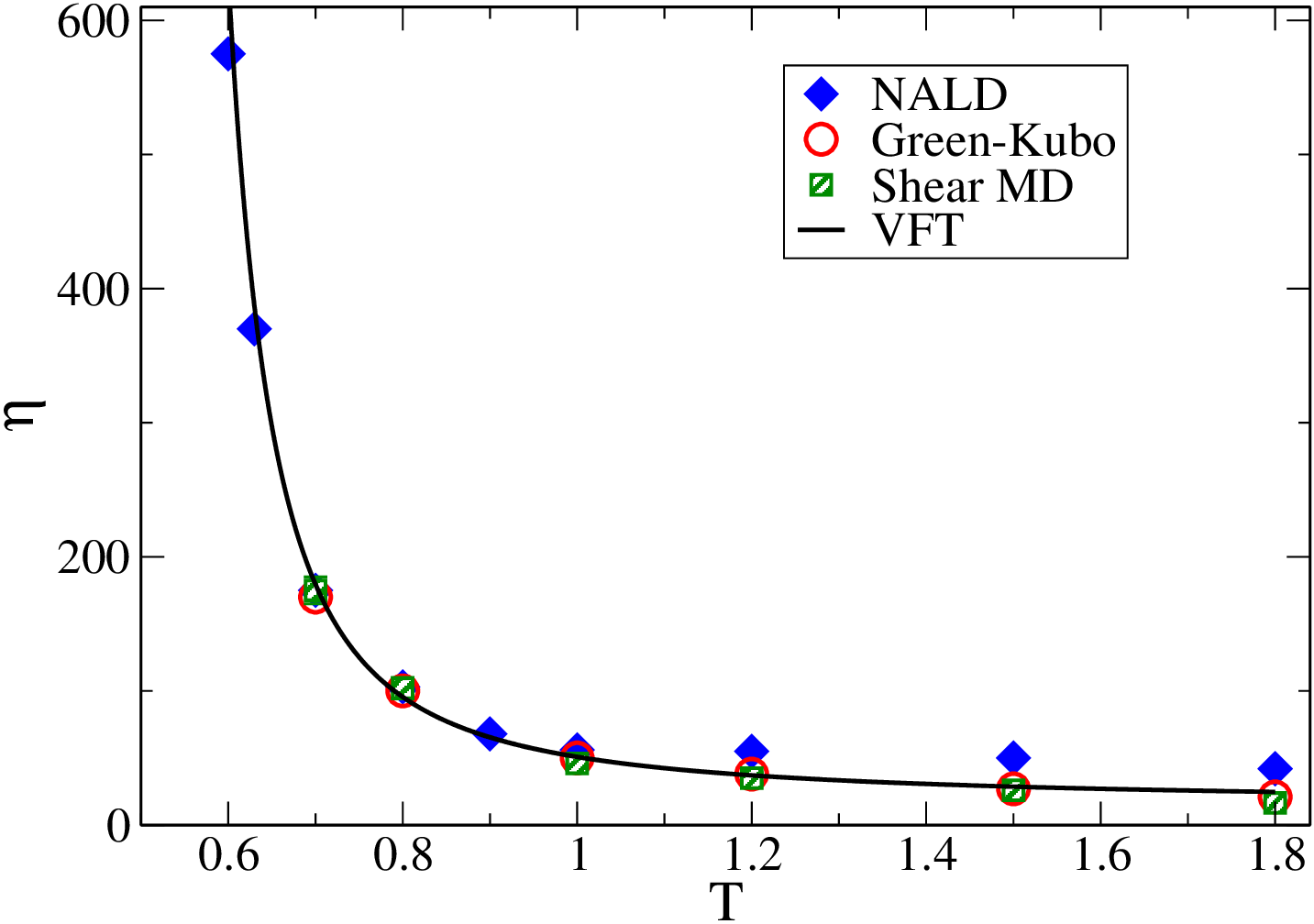}
    \caption{Comparison of temperature-dependent viscosity obtained using NALD (filled diamonds), Green-Kubo (open circles), and non-equilibrium molecular dynamics (filled squares) methods. The solid line represents the Vogel-Fulcher-Tammann (VFT) fit, confirming the glass transition temperature $T_{g} \simeq 0.4$. The quantities are in Lennard-Jones units.}
    \label{viscosity}
\end{figure}


In Fig.~\ref{viscosity_with_time}(b) we plot viscosity as a function of time at different temperatures. From the figure, we observe that the time-dependent viscosity saturates, making it reliable to extract the viscosity in the long-time plateau region dependent on temperature. At higher temperatures in the liquid state, the saturation region is reached quickly, whereas in the supercooled region with the decrease in temperature, it becomes increasingly difficult to achieve the long-time plateau. We have used these viscosity calculations as a reference to compare the viscosity from the theoretical NALD route.

Alternatively, viscosity results are obtained from non-equilibrium molecular dynamics simulations using the linear response relation; see Eq.~\ref{Shear_MD}. In Fig.~\ref{stress-strain}(a), we plot \(\eta = \sigma_{xy} / \dot{\gamma}\) as a function of shear strain \(\gamma\) at different temperatures with a fixed shear strain rate, where the shear strain rate is $10^{-3}$ for higher temperatures ($T = 1.5$ and $T = 1.0$) and $10^{-4}$ for lower temperatures ($T = 0.8$ and $T = 0.7$). The shear stress exhibits characteristic features: at lower temperatures, it initially increases linearly with shear strain up to a certain value of shear strain, after which a broader peak is observed. Then, the shear stress decreases and fluctuates around a mean value up to sufficiently large strain, eventually reaching a steady-state value. In contrast, at higher temperatures, shear stress increases linearly up to a certain shear strain and quickly reaches a steady state. The steady-state value in the plot represents the viscosity at the given shear strain rate and temperature.

In the supercooled regime, polymer melts exhibit significant shear thinning \cite{Datta_2023, Hara_2025}, characterized by a substantial reduction in viscosity ($\eta$) under high shear rates \cite{Bonn_2017, Mizuno_2024}. To avoid shear thinning effects across the entire temperature range studied, and in Fig.~\ref{stress-strain}(b), we plotted the viscosity as a function of shear strain rate at different temperatures. At higher temperatures, a shear strain rate of \(10^{-3}\) is sufficient to obtain a reliable viscosity estimate, where viscosity remains independent of the shear strain rate, thus characterizing the Newtonian regime. However, in the lower temperature regime, shear thinning is observed up to a shear strain rate of \(10^{-4}\). In this temperature regime, viscosity depends on the shear rate, exhibiting either shear thinning or shear thickening behavior, which is characteristic of the non-Newtonian regime. The plateau region of viscosity in Fig.~\ref{stress-strain}(b) at different temperatures were extracted and compared with other viscosity data in Fig.~\ref{viscosity}.

In Fig.~\ref{viscosity}, the viscosity is plotted as a function of temperature and compared with values obtained from different methods. Open circles represent the results from the Green-Kubo method, filled squares correspond to nonequilibrium MD (NEMD), and close diamonds are calculated using the NALD method (Eq.\ref{viscosity_eq}). The NALD method is particularly important for the lower temperature region near the glass transition temperature, where the viscosity exhibits a dramatic increase \cite{KSZ}. The solid line represents the Vogel-Fulcher-Tammann (VFT) fit \cite{Douglas_PANS_2009}, given by the equation $ \eta = \eta_0 \exp \left[ {A}/{(T - T_g)} \right]$ where $A = 0.75$ is a temperature-independent parameter, $\eta_0 = 15$, and \( T_g \simeq 0.4 \) is the glass transition temperature. The results show a very good agreement in the supercooled region, while some deviations are observed above the melting temperature. However, the overall trend is well captured across the entire temperature range, demonstrating the robustness and consistency of our approach.

\section{SUMMARY AND CONCLUSIONS}\label{Conclusion}
We investigated the microscopic origin of viscosity in model supercooled polymeric liquids from first principles using non-affine lattice dynamics (NALD). Using a model system of coarse-grained polymer melt (Kremer-Grest), we generated equilibrium configurations in the supercooled regime. These configurations were then used to calculate the vibrational density of states and affine force-field correlators \cite{Zaccone_book}, which were subsequently used to compute the complex shear modulus \cite{Zaccone_2023}. The viscosity of the system was directly obtained from the zero-frequency limit of the loss modulus. The temperature dependence of viscosity in the supercooled regime is accurately captured by this approach. For comparison, we also calculated the viscosity using the Green-Kubo (GK) method and non-equilibrium MD (NEMD) shear simulations over the accessible temperature range. By comparing the viscosity results obtained from different methods in both the liquid and supercooled regimes, we demonstrated that the NALD approach reliably predicts the viscosity, particularly in the supercooled regime closer to glass transition where experimental measurements and traditional simulation methods based on GK are challenging.


The NALD framework requires a molecular friction parameter $\nu$ as an input. In this study, we have used a Langevin thermostat in the MD simulations with a fixed dissipation timescale $\tau = 1$, which works as an effective inverse friction parameter for our system, and as such it was implemented in NALD, thus leaving no adjustable parameters. In a previous study by Palyulin {\it et al.} \cite{palyulin2018parameter} it was shown that the viscoelastic response in the NALD framework could be studied as a Markovian system where the friction of the system can be controlled using the Langevin thermostat. In general, the friction kernel can be derived using a particle-bath Hamiltonian framework by solving the Euler-Lagrange equations for the coupled dynamics of a tagged particle and heat bath oscillators, following Zwanzig’s formalism \cite{Zwanzig_1965, Zwanzig_1973,Zaccone_2023,Ceotto,Gottwald_2016}. In the NALD framework used thus far, the friction is a constant \cite{lemaitre_2006,palyulin2018parameter,vaibhav2024time,Elder}, because in coarse-grained systems it matches the value of the Langevin thermostat damping parameter. In the atomistic systems studied thus far, it is determined by matching the NALD calculation with the nonequilibrium molecular dynamics (NEMD) of the viscoelastic modulus at the high-frequency plateau \cite{Elder,vaibhav2024time}.
However, in experimental and real-world systems, the microscopic friction is typically time-dependent and non-Markovian \cite{Lickert_2020, Milster_2024,Netz_2019,Netz_2024}. Although the memory kernel can be extracted from MD simulations \cite{Tanja,Netz_2023}, a direct estimation based on the fluctuation-dissipation theorem \cite{Zwanzig_1965, Zwanzig_1973}, associated with the generalized Langevin equation, can be obtained from the time autocorrelation function of the stochastic force \cite{Jung_2017}. Another approach to measure the memory kernel involves computing the momentum autocorrelation and force correlation functions \cite{Gottwald_2016}. Through the analysis of memory kernels, it should be possible to recover the expected long-time transport behavior. Accurate evaluation of the memory kernel under favorable conditions can provide insights into the viscoelastic response, facilitating the design of functional materials for practical applications \cite{Milster_2024, vaibhav2024time}. Hence, in future studies, the friction parameter should be handled as a parameter that is derived from the microscopic input of the system. Furthermore, this framework can be extended to investigate how different frequency modes in the vibrational spectrum (vDOS) influence the viscoelastic modulus and transport properties in atomistic systems \cite{vaibhav2024time}, soft jammed amorphous \cite{Hara_2025}, colloidal glass \cite{Mason_1995, Negi_2014}, polymer nanocomposites \cite{Bonn} and granular fluids \cite{Cheng_2014}.
A crucial point for improvement in future work will be the scaling up of the eigenmodes computation towards lower frequencies, which is especially challenging for atomistic systems.

\section*{Acknowledgments}
AS, VV, and AZ gratefully acknowledge funding from the European Union through Horizon Europe ERC Grant number: 101043968 “Multimech”. VV acknowledges the computational resource provided via the project ``TPLAMECH'' on INDACO platform at the HPC facility of Università degli Studi di Milano. AZ gratefully acknowledges funding from the US Army DEVCOM Army Research Office through contract nr.  W911NF-22-2-0256.

\section*{AUTHOR DECLARATIONS}
\subsection*{Conflict of Interest}
The authors have no conflicts to disclose.

\section*{Data Availability}
The data that support the findings of this study are available from the corresponding author upon reasonable request.

\bibliographystyle{apsrev4-1}
\section*{References}
\bibliography{article}

\end{document}